\documentclass[conference,compsoc]{IEEEtran}
\usepackage[nocompress]{cite}
\usepackage[caption=false,font=footnotesize,labelfont=sf,textfont=sf]{subfig}
\usepackage{fixltx2e}
\usepackage{stfloats}

\usepackage{amsmath,amssymb,amsfonts}
\usepackage{algorithmic}
\usepackage[pdftex]{graphicx}
\usepackage{textcomp}
\usepackage{xcolor}
\usepackage{booktabs}
\usepackage{comment}
\usepackage[noend]{distribalgo}
\usepackage{colortbl}
\DeclareGraphicsExtensions{.pdf,.jpeg,.png}
\usepackage{array}
\usepackage[caption=false,font=footnotesize,labelfont=sf,textfont=sf]{subfig}

\usepackage{stfloats}
\usepackage{url}

\begin{document}
%
\title{Just-in-Time Aggregation for Federated Learning}

\author{\IEEEauthorblockN{K. R. Jayaram}
\IEEEauthorblockA{IBM Research AI, USA}
\and
\IEEEauthorblockN{Ashish Verma}
\IEEEauthorblockA{IBM Research AI, USA}
\and
\IEEEauthorblockN{Gegi Thomas}
\IEEEauthorblockA{IBM Research AI, USA}
\and
\IEEEauthorblockN{Vinod Muthusamy}
\IEEEauthorblockA{IBM Research AI, USA}
}

\maketitle

\begin{abstract}
The increasing number and scale of federated learning (FL) jobs necessitates
resource efficient scheduling and management of aggregation to make the economics of
cloud-hosted aggregation work. Existing FL research has focused on the design of FL algorithms and optimization, and less
on the efficacy of aggregation. Existing FL platforms 
often employ aggregators that actively wait for model updates. This wastes computational
resources on the cloud, especially in large scale FL settings where parties are intermittently
available for training. 

In this paper, we propose a new FL aggregation paradigm -- ``just-in-time'' (JIT) aggregation
that leverages unique properties of FL jobs, especially the periodicity of model updates,
to defer aggregation as much as possible and free compute resources for other FL jobs or other
datacenter workloads. We describe a novel way to prioritize FL jobs for aggregation, and demonstrate
using multiple datasets, models and FL aggregation algorithms 
that our techniques can reduce resource usage by 60+\% when compared to eager aggregation used
in existing FL platforms. We also demonstrate that using JIT aggregation has negligible overhead
and impact on the latency of the FL job.
\end{abstract}


\section{Introduction}~\label{sec:intro}

\begin{figure}[ht]
    \centering
    \includegraphics[width=\columnwidth]{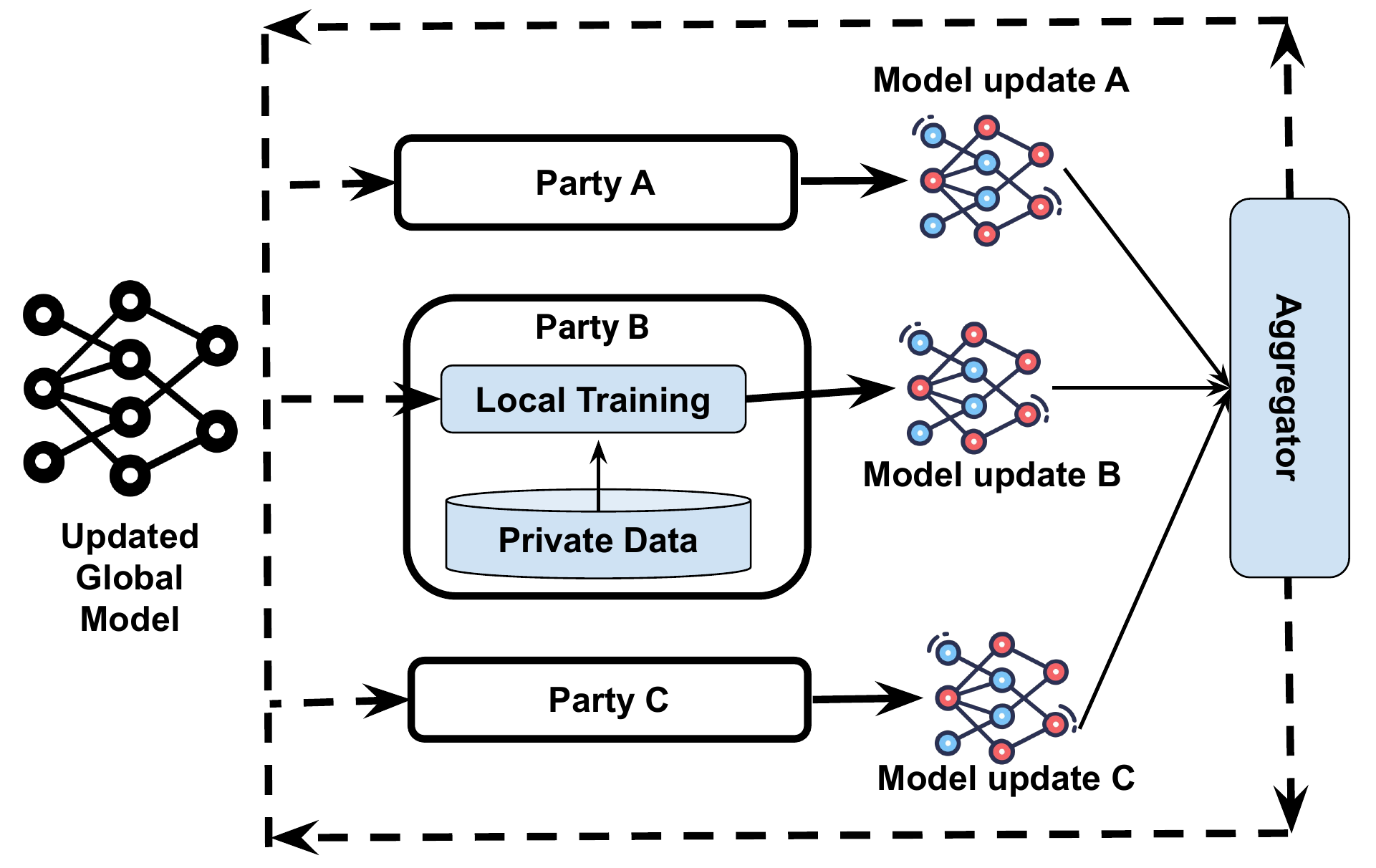}
    \caption{Federated Learning illustration with three parties}
    \label{fig:floverview}
\end{figure}

Federated learning (FL)~\cite{kairouz2019advances}, illustrated in Figure~\ref{fig:floverview}, 
is a type of machine learning which 
avoids centralization of data, and enables participants to train models without sharing their 
private data with cloud services and with each other.
At the start of and FL job, parties agree among themselves 
on the architecture of the machine learning model (e.g., the specific neural network) to be trained
and hyperparameters to be used and train locally (i.e., within their controlled domains). 
Only model updates are shared,
typically, to a central aggregator server hosted by a cloud service provider. The aggregator fuses
local model updates from parties to compute a global aggregated model which is 
communicated back to the parties. Providing strong privacy protection to participant data is a 
key goal of FL and central to its definition.

FL is \emph{typically} deployed in two scenarios: \emph{cross-device} and \emph{cross-silo}.
The cross-device scenario involves a large number of parties ($>1000$), but each party has a small 
number of data items, constrained compute capability, and limited energy reserve 
(e.g., mobile phones or IoT devices). 
They are highly unreliable and intermittently available, i.e, 
expected to drop and rejoin frequently. Examples include a large 
organization learning from data stored on employees' devices and a device manufacturer training a model 
from private data located on millions of its devices (e.g., Google Gboard~\cite{bonawitz2019towards}). 
Contrarily, in the cross-silo scenario, the number of parties is small, but each party has extensive 
compute capabilities (with stable access to electric power and/or equipped with hardware accelerators) 
and large amounts of data. There is reliable participation throughout the entire federated 
learning training life-cycle. Examples include 
multiple hospitals collaborating to train a model from radiographs (e.g., NVIDIA's work on COVID
CT scans~\cite{nvidia-covid}), multiple banks 
collaborating to train a credit card fraud detection model, etc.


FL has been shown to achieve significant increases 
in model utility when compared to parties training solely on their local datasets. 
Increasing adoption of FL has, in turn, increased the need for 
FL-as-a-service offerings by public cloud providers, which serve as a nexus 
for parties in an FL job and aggregate/fuse model updates 
(e.g., IBM Federated Learning (IBMFL)~\cite{ibmflpublic, ibmfl}). 
Such cloud services have to scale effectively to support multiple concurrent FL jobs and 
multi-tenancy. 
Hence, effective \emph{aggregation of model updates} is a key problem in FL,
 when viewed from either a performance, scalability, resource efficiency/cost, or
 privacy perspective. However, these aspects remain under-addressed in FL research.

 Performing aggregation in a resource and cost effective manner, while scaling to a large
number of participants, is challenging
for cloud service providers. The key question 
for a cloud/datacenter-based FL service becomes when to schedule aggregation without delaying an FL
job, while also ensuring high datacenter resource utilization. A related question which also poses a challenge is how
long to keep the aggregator deployed waiting for model updates. This is primarily due to two factors -- 
the intermittent availability of parties and heterogeneity of their hardware and data.

This paper revisits the traditional ``always-on''  aggregation 
paradigm in FL and makes the following technical contributions:

\begin{enumerate}

\item A detailed description of why efficient aggregation of model updates is a hard problem, especially when viewed from a cloud or service provider's perspective.

\item An illustration of how unique properties of FL jobs, namely periodicity and linearity, can be used to 
aggregate efficiently.

\item A new ''just-in-time'' (JIT) aggregation strategy for FL jobs that defers aggregation as much as possible to
free compute resources for other FL jobs or other datacenter workloads.

\item An empirical comparison of lazy aggregation against eager and batched aggregation using
    three different models/datasets and two aggregation algorithms to demonstrate that it does not increase the latency of FL jobs but leads to significant resource savings.
\end{enumerate}

\section{Background}

\subsection{FL Jobs and Model Aggregation}


An FL job involves parties performing local training on their data, 
sharing the weights of their model (also called a \emph{model update}) with the aggregator, 
which aggregates the weight vectors of all parties using a fusion algorithm.
A model update (whether weight update or gradient update) is flattened, and represented as a list of
one-dimensional vectors (e.g., in Tensorflow/Keras), with each vector corresponding to a layer.
Ways to fuse/aggregate these model updates involve
coordinate-wise computations on these vectors (averaging, weighted averaging, multiplication etc.).
That is, aggregation $\oplus$ of two model updates $M_1[1,\ldots,n]$ and $M_2[1,\ldots,n]$
involves applying a function $f$ to each component/element of the update vector
$M_1 \oplus M2 ~=~ [f(M_1[1],M_2[1]),\ldots, f(M_1[n],M_2[n])]$. 
Then, the merged/aggregated model is sent back to all
parties for the next round of training on their local datasets. 

Like regular (centralized) machine learning training which makes several passes over a 
centralized dataset, an FL job proceeds over a number of model fusion/synchronization rounds, determined 
by the batch size ($B$) used. While model fusion after every minibatch ($B$) is possible, typically
parties in an FL job synchronize every local epoch, i.e., they train by making a pass over their entire
local data set before fusing local models. The performance of an FL job has two dimensions 
(i) Utility -- is the accuracy of the federated model much better than that of the locally trained model,
and (ii) Latency, which includes training and aggregation latency. Training latency depends on the amount
of data at each party and the hardware available for training. \emph{Aggregation Latency} is the time taken
for aggregation to complete after the last required model update is available. Aggregation latency is the manner
in which the effectiveness and performance of the aggregator is perceived by the parties; a lower
aggregation latency is better.

\subsection{Active vs. Intermittent Participants}


\emph{Active} participation means that parties have dedicated resources to the FL job, and will
    promptly respond to aggregator messages. That is, for every synchronization round, once the aggregator sends
    the updated model, the party starts the next local training round and sends a (local) model update as soon as 
    training is done. Active participation does not mean specific types of optimization algorithms are used. Generally, active participation is only seen in small scale FL jobs, and more often in cross-silo settings.
    
\emph{Intermittent} participation means that for every FL round, each party trains at its convenience, 
or feasibility. This may be when connected to power in the case of mobile phones, tablets and laptops;
when (local) resource utilization from other computations is low and when there are no pending jobs with higher priority. In these scenarios, the aggregator expects to hear from the parties \emph{eventually} - typically over several minutes or hours and sometimes once a day in the case of mobile phones. Large-scale FL jobs almost always involve intermittent parties -- at scale, it is unreasonable to expect that all parties participate at the same pace.

\subsection{Heterogeneity}

In FL, party heterogeneity takes two forms -- 
(i) parties with varying compute capabilities and (ii) parties having different amounts of data.
Although heterogeneity is intuitive and easy to visualize in cross-device settings, our experience 
has been that it affects cross-silo settings as well, both with active and intermittent parties.
Some examples we have observed are: hospitals of different sizes in different timezones actively participating 
in an FL job from their datacenters (data and compute heterogeneity), an FL job consisting of different types
of devices (laptops, mobile phones, desktops, etc.). 

The arguments regarding minimizing aggregation latency apply here as well. If an FL job merges model updates
per local epoch, the amount of data at each party determines when the local training completes and model
update arrives. Training time is also dependent on the compute capabilities of each party.
Party heterogeneity, thus increases the intermittent nature of model updates, making aggregation challenging.

\section{Design/Deployment Choices for Aggregation}

In this section, we describe four aggregation strategies --
\emph{Eager Always-on, Lazy, Eager Serverless, and JIT Serverless},
examine their pros/cons and implementation challenges.

\begin{figure*}[htb]
    \centering
    \includegraphics[width=0.7\textwidth]{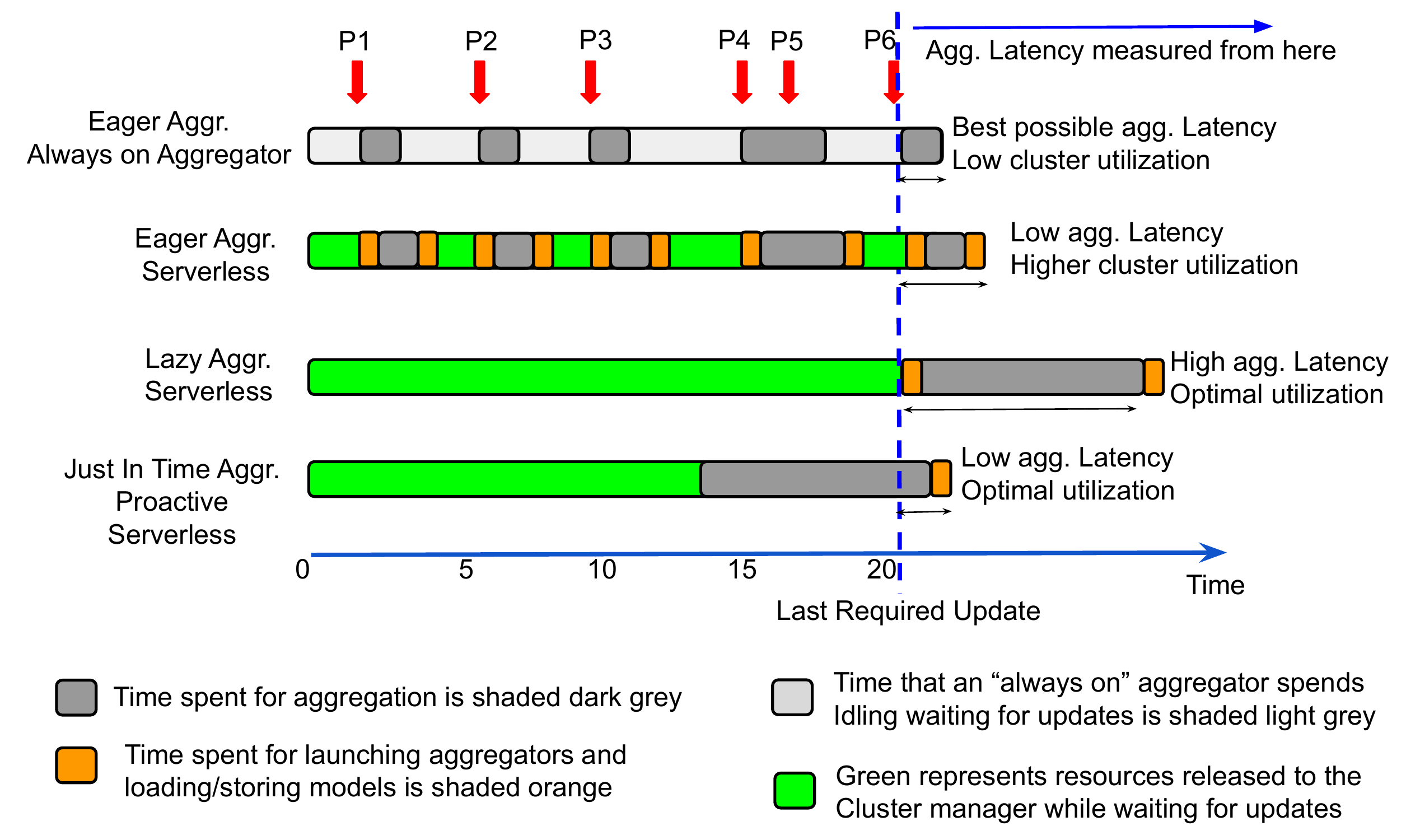}
    \caption{Aggregation Design Options}
    \label{fig:agg_choices}
\end{figure*}

{\bf Eager Always-on:} 
In this strategy, used in FL platforms like IBM FL~\cite{ibmfl, ibmflpublic},
FATE~\cite{fate}, NVFLARE~\cite{nvflare}, aggregators, either deployed as 
servers, virtual machines or containers are 
``always on'', i.e., deployed continuously throughout the FL job,
waiting for updates, and handle each update as soon as it arrives.
This is illustrated in Figure~\ref{fig:agg_choices} with the dark grey representing
aggregation and light grey representing periods when the aggregator is idle.
Assume an FL job with six parties ($P1 - P6$) that proceeds
over several model fusion rounds. Assume that round $r$ starts at time $t=0$;
the parties send their model updates intermittently over 20s and that
aggregation takes $1s$ for a pair of model updates.  
Eager aggregation is completed at time $t=21$, 
because the aggregator 
handles the updates from $P1 - P5$ while waiting for $P6$, and immediately handles $P6$'s update at $t=20$.
However, this requires the aggregator to be scheduled and provisioned for 21 time units, while aggregation only 
takes 6 time units, thereby idling for $\frac{6}{21}$ or 71.4\% of the time. The only benefit is that the \emph{aggregation latency} -- the time taken for aggregation to complete
after the last model update arrives, is minimal. 

{\bf Lazy Dynamic/Serverless:}
The Lazy strategy, as the name implies, schedules the aggregator for all updates
only after the last update arrives. This is optimal from a cluster utilization
point of view but can result in high aggregation latency (Figure~\ref{fig:agg_choices}). 
Lazy serverless can be useful
when there only a few parties, but aggregation latency grows quickly as the number of parties
increases. For some FL jobs, aggregation can dominate training when the lazy serverless strategy
is used.

{\bf Eager Dynamic/Serverless:}
One way to improve eager aggregation is to deploy the aggregator dynamically every time
a model update arrives. This can be done either using technologies like Kubernetes pods
or through serverless (cloud) functions. When compared to an ``always-on'' strategy,
Eager serverless has overheads at deployment and shutdown time. At deployment time, there is
the overhead of scheduling the serverless function and the time taken to load aggregator state
from stable datacenter/cloud storage. At the end of each deployment, the aggregated model
and other state has to be checkpointed back to stable storage. These overheads are
illustrated in Fig.~\ref{fig:agg_choices} (orange color). Eager serverless improves 
cluster utilization when compared to the ``always-on'' deployment, because it relinquishes
cluster resources more frequently (green color in Figure~\ref{fig:agg_choices}). Overheads increase 
aggregation latency somewhat, but it is still low compared to using a lazy aggregation strategy.
Using any dynamic deployment strategy (including serverless functions) requires that
model updates be buffered somewhere in the datacenter, e.g., a message queue like Kafka or a cloud object store.


{\bf JIT Aggregation:}
Our goal in this paper, is to design and implement a ``Just In Time'' aggregation strategy
as illustrated in Fig.~\ref{fig:agg_choices} -- a strategy that starts aggregation just
in time anticipating the arrival of the last model update; a strategy that optimizes cluster 
utilization and aggregation latency. The key question to answer to achieve this is to 
schedule aggregation at just the ``right time'' and anticipate when a model update is going to
arrive. We describe this in the next section

\section{Predicting the Next Update}

\begin{figure}[htb]
    \centering
    \includegraphics[width=0.8\columnwidth]{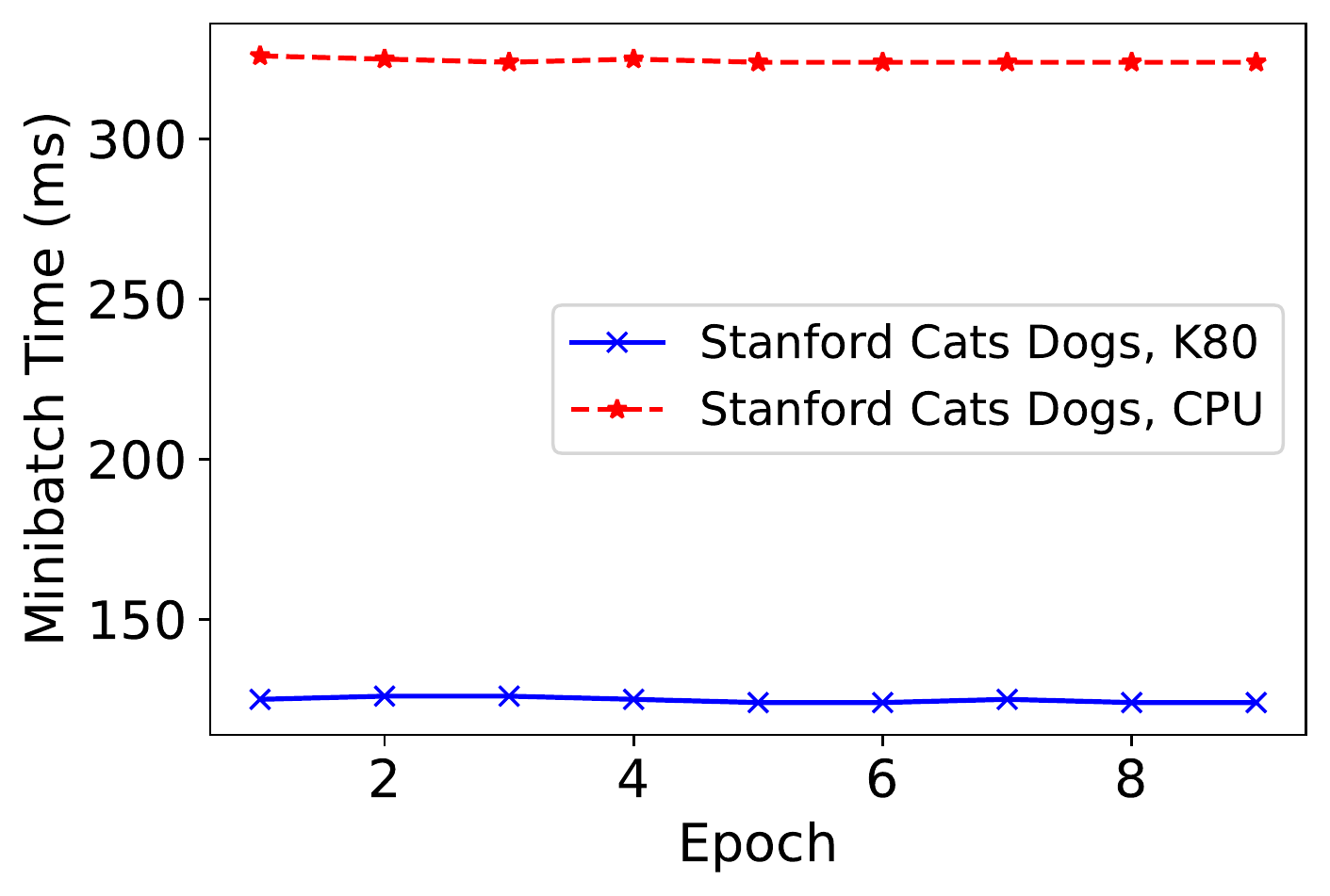}
    \includegraphics[width=0.8\columnwidth]{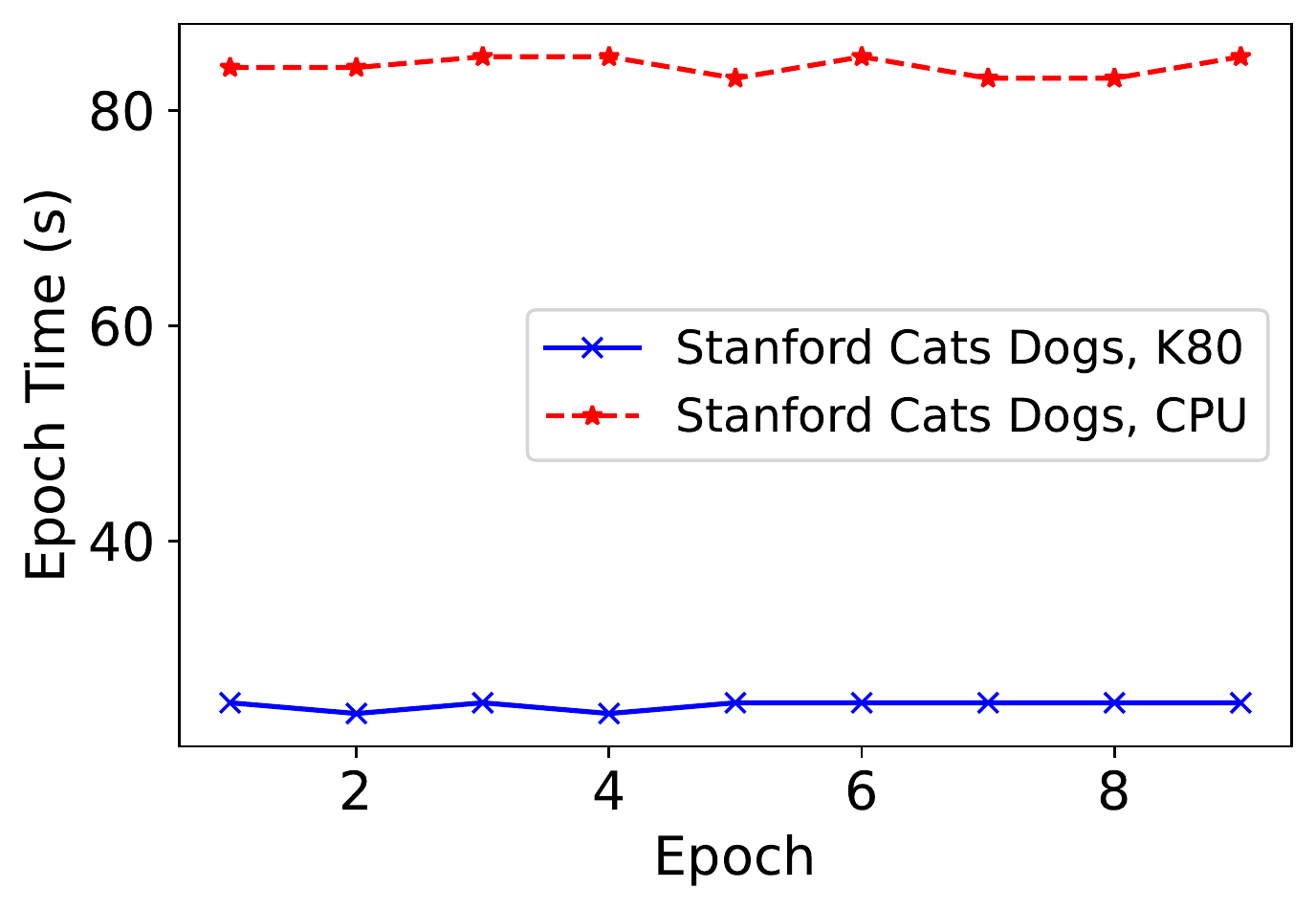}

    \caption{Minibatch time (left) and Epoch time (right) remain fairly constant across epochs in the absence of changes to data and hardware. Efficientnet-B7 on Stanford Cats/Dogs}
    \label{fig:constancy_mb}
\end{figure}

\begin{figure}[htb]
    \centering
    \includegraphics[width=0.8\columnwidth]{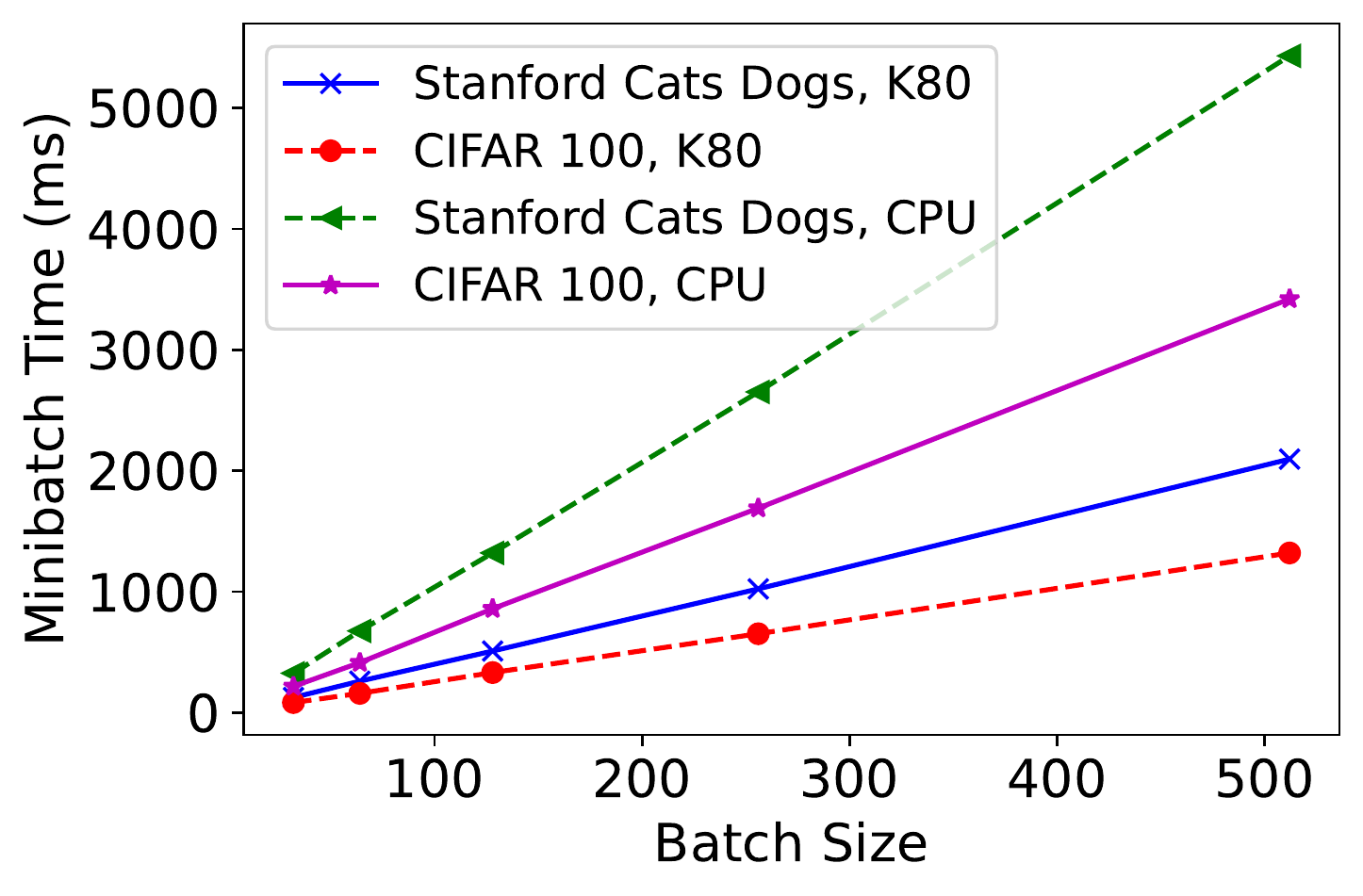}
    \includegraphics[width=0.8\columnwidth]{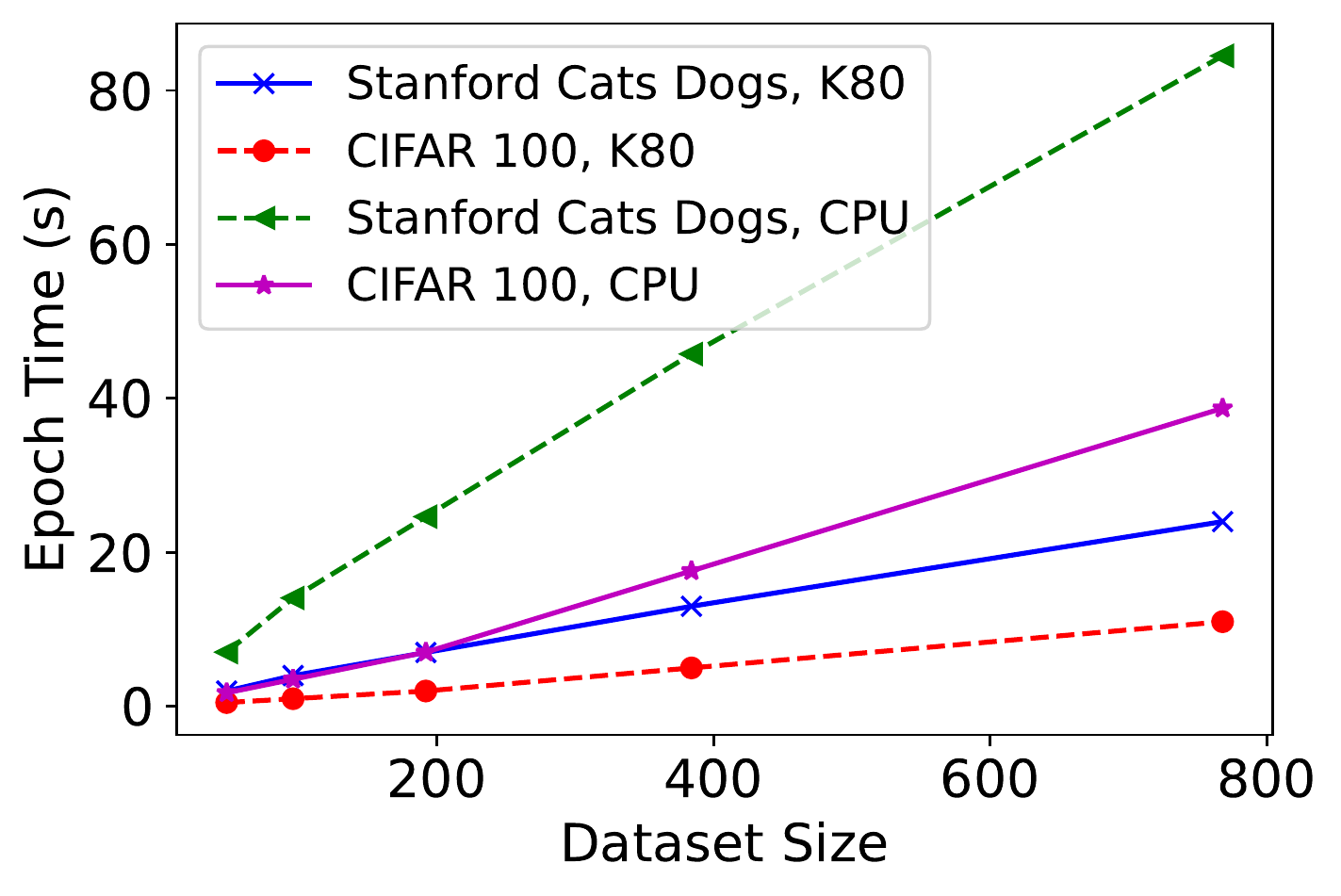}
    \caption{Minibatch time vs. batch size and Epoch time vs. Dataset size}
    \label{fig:mb_bs}    \label{fig:ds_et}
\end{figure}



To do any form of JIT aggregation, predicting the arrival of the next model update
becomes vital. In this paper, we leverage two key properties of many machine learning
workloads -- \emph{Periodicity} and \emph{Linearity} to make educated guesses about when
the next model update is going to arrive (or not arrive).


\subsection{Periodicity of Model Updates \& Active Parties}


The ``local'' part of FL is similar to traditional ML, i.e., model training makes
several passes over the dataset (each pass is called an epoch). 
A local model update is generated either once every epoch
or for every batch of data items processed (also called mini-batch in gradient descent 
terminology). From our experience building and operating machine learning platforms, we have 
observed that minibatch times and epoch times are constant if the training dataset at a party does not
change between epochs and if there are no competing workloads. We have validated these observations
with multiple experiments, one of which is illustrated in Figure~\ref{fig:constancy_mb}.
Here, we train the EfficientNet-B7 neural network model on the Stanford Cats/Dogs and CIFAR100 datasets on different hardware -- NVIDIA K80 GPU and an eight core (Core i9) Intel CPU using Tensorflow.
This isn't surprising, since each minibatch involves the same number of data items, 
and since the data items are
normalized (e.g., images converted to the same resolution), each minibatch, and consequently
each epoch takes roughly the same time on the same hardware in the absence of competing workloads.
ML engineers have observed this behavior across a variety of models/datasets.
Consequently, from an FL aggregation perspective, an active participant 
should roughly take the same 
time for each FL training round, making model updates from a given party \emph{periodic}.

\subsection{Linearity}

Furthermore, in Figure~\ref{fig:mb_bs}, we observe a linear relationship
between the minibatch time and the batch size, as well as a linear relationship between
the epoch time and the dataset size. We observe this for two different datasets -- Stanford Cats/Dogs
and CIFAR100 with EfficientNet-B7; the behavior is similar on other datasets.
Again, this behavior is intuitive -- the time taken to train a neural network
with 32 images (batch size of 32) will roughly be twice the time taken to train with 
16. Similarly, time taken to complete a local epoch with 32 GB of local data will be 
roughly twice that with 16 GB of local data. Due to this linearity, even when training data changes (e.g. new data items are added
or some data is lost), linear regression can be used to predict new epoch times from previous
measurements. Periodicity and linearity can be very useful in 
predicting \emph{when} the next model update is going to be sent from a party, which in turn
can be used to determine when aggregation must be scheduled to meet SLA and efficiency requirements.

\subsection{Active vs. Intermittent Participation}\label{sec:waitingperiod}

Active parties dedicate resources to model training, which means that
they send periodic model updates every $t_{train}+t_{comm}$, where
$t_{train}$ is the time taken for local training and $t_{comm}$ is the 
time taken for transmitting the model to and from the aggregator.

With intermittent parties, typically, there is an agreement among parties and the aggregator
that parties train at their convenience, but this is not open ended and there are
timeouts.
Generally, for every FL round, each party is expected to respond 
to the aggregator within a certain time period $t_{wait}$ from the start of the 
FL round. Beyond this, the model update is ignored. $t_{wait}$
is highly application dependent (it can be minutes or hours or as long as a day) and
agreed at the start of the FL job by mutual consent. This also sets SLA expectations
for parties with respect to aggregation -- parties expect a new round to start
every $t_{wait}$ and expect aggregation to complete before that.
There is typically no incentive to complete earlier because parties may not
be able to start the next round (because e.g., the goal
may be to train every night or on data received during the day).
Our approach leverages these expectations to ensure that
aggregation is always complete within $t_{wait}$. 


\section{JIT Aggregation -- Design and Implementation}

\begin{figure}[h]
\centering
\includegraphics[width=\columnwidth, keepaspectratio=true]{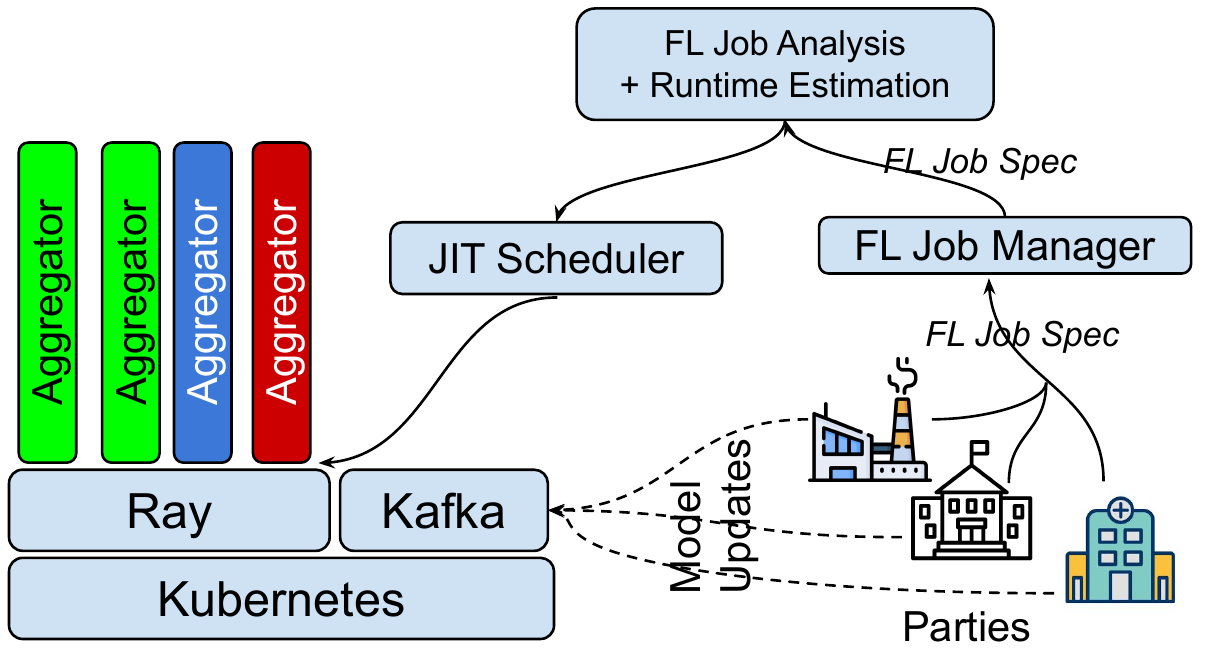}
\caption{Architectural Overview of JIT Aggregation}~\label{fig:archoverview}
\vspace{-0.7cm}
\end{figure}

Our core contribution lies in using training time estimation of machine learning
jobs to schedule aggregation in a resource efficient manner in federated learning settings,
while minimizing aggregation latency. An architectural overview of 
JIT aggregation in a cloud hosted FL platform 
is presented in Figure~\ref{fig:archoverview}, and the high level pseudocode of
the JIT aggregation is presented in Figure~\ref{algo:jitaggregation}.

\subsection{FL Jobs and Specs}

In existing FL systems~\cite{ibmfl, ibmflpublic}, the aggregator of every FL job knows the model architecture,
the optimizer/aggregation algorithm and the hyperparameters of the job.
Hyperparameters include learning rate, batch size and frequency of synchronization, i.e., the
frequency of global model update. Frequency of synchronization is typically once per local epoch,
but can also be once every few minibatches.
Agreement on the model architecture and hyperparameters is essential to set up the job. 
Other inputs specific to the FL job include $t_{wait}$ in the case of intermittent parties,
and the minimum number of parties that are needed (quorum) for an FL round to be successful.
Parties agree on these inputs and send an ``FL Job Specification'' to the aggregator (typically
a cloud service provider that hosts aggregation). Our system analyzes this specification
to predict the arrival of updates to schedule aggregation.

\subsection{Additional Input Needed From Parties}

Our system needs the following \emph{additional} information from parties in an FL job:
(i) \emph{mode of participation}, i.e., whether the party intends to participate \emph{actively},
(ii) \emph{training time} -- epoch time, minibatch time and size of the party's dataset or
\emph{party hardware information} -- number and type of CPUs/GPUs used for training and 
(iii) \emph{network bandwidth} between the party and aggregator.
Mode of participation is easy to provide. To estimate when the next update is likely 
to arrive, our technique relies on parties to directly provide local minibatch or epoch time --
these are measured by default by most machine learning frameworks including 
Tensorflow and PyTorch, without the programmer even having to write additional code.
If for some reason, parties are unwilling to provide these, there is the option of 
providing information about the hardware used for training from which minibatch time
is estimated using linear regression. For network bandwidth, we have implemented an 
extension to Tensorflow using standard Linux tools
to periodically measure average network bandwidth between the party and the aggregator.
From periodic measurements, we compute $\mathcal{B}_d$ and $\mathcal{B}_u$, which are
the average aggregator $\rightarrow$ party and party $\rightarrow$ aggregator bandwidths
respectively. The frequency of measurement can be configured depending on the party (sensor vs. mobile phone vs. datacenter).
Information about the job including all the above inputs are stored in a persistent store like 
MongoDB.

\newcommand{\arrival}[1]{\textsc{arrival}(#1)}
\newcommand{\commtime}[1]{\ensuremath{t_{comm}^{(#1)}}}
\newcommand{\traintime}[1]{\ensuremath{t_{train}^{(#1)}}}
\newcommand{\aggtime}[1]{\ensuremath{t_{agg}^{#1}}}
\newcommand{\waittime}[1]{\ensuremath{t_{wait}^{#1}}}
\newcommand{\updtime}[1]{\ensuremath{t_{upd}^{(#1)}}}
\newcommand{\updtimemax}[1]{\ensuremath{t_{rnd}^{#1}}}
\newcommand{\epochtime}[1]{\ensuremath{t_{ep}^{(#1)}}}
\newcommand{\mbtime}[1]{\ensuremath{t_{mb}^{(#1)}}}
\newcommand{\inlineif}[1]{\textbf{if}~#1}
\newcommand{\Job}[1]{\mathcal{J}_{#1}}
\newcommand{\Party}[1]{\mathcal{P}_{#1}}
\newcommand{\ms}[1]{\mathcal{M}}
\newcommand{\bwdn}[1]{\mathcal{B}_{d}^{(#1)}}
\newcommand{\bwup}[1]{\mathcal{B}_{u}^{(#1)}}
\newcommand{\bwdc}[1]{\mathcal{B}_{dc}^{#1}}
\newcommand{\priority}[1]{\mathcal{P}^{#1}}
\newcommand{\deadline}[1]{\mathcal{D}^{#1}}
\newcommand{\roundstart}[1]{\textsc{start\_round(#1)}}
\newcommand{\aggfreq}[1]{\ensuremath{f_{agg}}}
\newcommand{\aggjobdef}[1]{\mathcal{A}^{#1}}

\begin{figure}[htb]
\begin{distribalgo}[1]
\UPON{\arrival{FLJob $\Job{}$}}
    \STATE{$\aggfreq{} \leftarrow \textsc{get\_agg\_frequency}(\Job{})$}
    \STATE{$\waittime{} \leftarrow \textsc{get\_wait\_time}(\Job{})$}
    \STATE{$\ms{} \leftarrow \textsc{get\_model\_size}(\Job{})$}
    \STATE{$\{\Party{1},\ldots,\Party{N}\} \leftarrow \textsc{get\_parties}(\Job{})$}
    \FORALL{$\Party{i} \in \{\Party{1},\ldots,\Party{N}\}$} 
        \STATE{$\traintime{i} \leftarrow \begin{cases}
                                        \waittime{} ~~ \inlineif{{\Party{i}} \mbox{ intermittent} \\
                                        \epochtime{i} ~~ \inlineif{\aggfreq{} = 1} \mbox{ local epoch}} \\
                                        N_{mb} \times \mbtime{i} ~~ \inlineif{\aggfreq{} = N_{mb}} \mbox{ minibatches} \\
                                        \end{cases}$}
        \STATE{$(\bwdn{i}, \bwup{i}) \leftarrow \textsc{get\_bandwidth}(\Party{i})$}                          
        \STATE{$\commtime{i} \leftarrow \ms{}/\bwdn{i} + \ms{}/\bwup{i}$} \COMMENT{Time spent transfering models}
        \STATE{$\updtime{i} \leftarrow \traintime{i} + \commtime{i}$} \COMMENT{When is $\Party{i}$ going to update?}

        \ENDFOR
    \STATE{$\updtimemax{} \leftarrow max(\{\updtime{1},\ldots,\updtime{N}\})$} \COMMENT{Estimated time for each round}
    \STATE{$\textsc{fljobs}[\Job{}] \leftarrow \{\updtimemax{},\aggtime{}\}$} \COMMENT{Store estimated parameters}
    \STATE{$\aggtime{} \leftarrow \frac{N_{parties} \times t_{pair}}{C_{agg} \times N_{agg}} + \frac{\ms{k}}{\bwdc{}}$} \COMMENT{Est. aggregation time. Section~\ref{sec:aggtime}}
    
\ENDUPON

\UPON{\roundstart{$\Job{}$}}
\STATE{$\aggjobdef{} \leftarrow \textsc{create\_aggregators}(\Job{})$} \COMMENT{Create aggregator tasks}
\STATE{$\{\updtimemax{},\aggtime{}\} \leftarrow \textsc{fljobs}[\Job{}]$}
\STATE{$\textsc{set\_priority}(\aggjobdef{}, \updtimemax{} - \aggtime{})$} \COMMENT{Section~\ref{sec:priority}}
\STATE{$\textsc{set\_timer}(\aggjobdef{},\updtimemax{} - \aggtime{})$} \COMMENT{Section~\ref{sec:priority}}
\ENDUPON

\UPON{$\textsc{timer\_alert}(\aggjobdef{})$}
\IF{$\aggjobdef{}$~not~executing} 
\STATE{$\textsc{force\_trigger}(\aggjobdef{})$} \COMMENT{Deadline reached. Section~\ref{sec:priority}}
\ENDIF
\ENDUPON
\end{distribalgo}

\caption{High-level Pseudocode of JIT Aggregation Scheduler}~\label{algo:jitaggregation}
\end{figure}

\subsection{Local Training Time Estimation}

For each active party in an FL job $\Job{}$,  the expected time for a party $\Party{i}$ 
to finish local training $\traintime{i}$ is estimated (Fig.~\ref{algo:jitaggregation}, Line 7) as:

\begin{itemize}
    \item $\epochtime{i}$, if $\epochtime{i}$ is provided by the (active) party and the models
    are aggregated once per local epoch.
    \item $\mbtime{i} \times N_{mb}^{k}$ if $\mbtime{i}$ is provided by the (active) party $\Party{i}$
    and the model fusion for job $\Job{k}$ happens every $N_{mb}^{k}$. If $\mbtime{i}$ is not provided by the party,
    it can be estimated using linear regression if the hardware and memory available to the party are known. 
    \item $\waittime{k}$ of job $\Job{k}$ if the party is intermittent. 
\end{itemize}

At the start of each FL round, the party downloads a global model to use for training, and at the end of the
local training, it uploads the model update to the aggregator. The time taken for this $\commtime{i}$ is therefore
$\frac{model\_size}{\bwdn{i}} + \frac{model\_size}{\bwup{i}}$ (Fig.~\ref{algo:jitaggregation}, Line 9). Hence, the model update from $\Party{i}$ can be
expected to arrive at $\updtime{i}~=~\traintime{i}+\commtime{i}$ (Fig.~\ref{algo:jitaggregation}, Line 10).

\subsection{Aggregation Time Estimation}~\label{sec:aggtime}

Each party in an FL job trains the same model; model updates merely differ in the values
assigned to the weights in the model. Hence, if the time taken at the aggregator to fuse a pair
of updates is $t_{pair}$, then the computation time taken to aggregate all updates from $N_{parties}$ is 
$t_{agg} = N_{parties} \times t_{pair}$. If model updates can be aggregated in parallel
(i.e., if the aggregation function is data parallel), and if $N_{agg}$ aggregator nodes (VMs/containers)
are used with each aggregator having $C_{agg}$ usable CPU/GPU cores, then the computation time taken
to complete aggregation is $(N_{parties}\times t_{pair})/(C_{agg} \times N_{agg})$.
$t_{pair}$ on a single CPU core/GPU can be easily computed offline on the aggregator before the FL job starts --
by randomly generating model updates (assigning random values to weights in the model) and 
measuring the time taken to fuse pairs of these randomly generated model updates. 
We also note that $C_{agg}$ is the number of \emph{usable} cores for aggregation -- for
CPU based aggregation, this is often equal to the number of CPU cores in the aggregator
node (VM or container). But, for GPU based aggregation, the number of available GPU cores
may be much higher than the number of model updates that can fit into GPU memory.
To the computation time, we add the communication time for loading models from the message queue to computation
time to obtain $\aggtime{}$ (Fig.~\ref{algo:jitaggregation}, Line 13, where $\bwdc{}$ is the intra-datacenter 
bandwidth).

\subsection{JIT Aggregation with Deadlines and Priorities}~\label{sec:priority}

Consider an FL job with $N$ parties, with the estimated model update times of $\{\updtime{1},\ldots,\updtime{N}\}$,
and estimated aggregation time $\aggtime{}$. This aggregation can be safely deferred from
the start of an FL round until $\updtimemax{} - \aggtime{}$ where $\updtimemax{} = max(\{\updtime{1},\ldots,\updtime{N}\})$.
This is because the goal of JIT aggregation is to minimize aggregation latency, which is the time taken
to complete aggregation after $\updtimemax{}$; consequently, aggregation should complete
soon after $\updtimemax{}$. Starting aggregation any time after $\updtimemax{} - \aggtime{}$
increases the probability of a higher aggregation latency.
Hence, we employ a timer to ensure that aggregation starts at $\updtimemax{} - \aggtime{}$.
This is the purest form of JIT aggregation.

But, we would like to be opportunistic (``greedy'')
and use the cluster if it is idle. Hence we combine timers with priorities. 
We set the priority of the aggregation task to $\updtimemax{} - \aggtime{}$ as well; a smaller
priority value indicates a higher priority job. Hence, if the Kubernetes cluster has idle
cycles before ($\updtimemax{} - \aggtime{}$), aggregation jobs are automatically scheduled
by the JIT scheduler according to their priority, and execute if there are model updates waiting in the 
message queue. Scheduling decisions are made every $\delta$ seconds, which is configurable.
If higher priority FL aggregation tasks or ther workloads arrive, lower priority aggregators
are pre-empted by checkpointing partially aggregated model updates using the message queue.
If there are no pending updates to aggregate, the JIT scheduler defers aggregation tasks, while
retaining their priority.

\section{Evaluation}
\label{sec:eval}

\begin{figure*}[htb]
    \centering
    \includegraphics[width=0.32\textwidth]{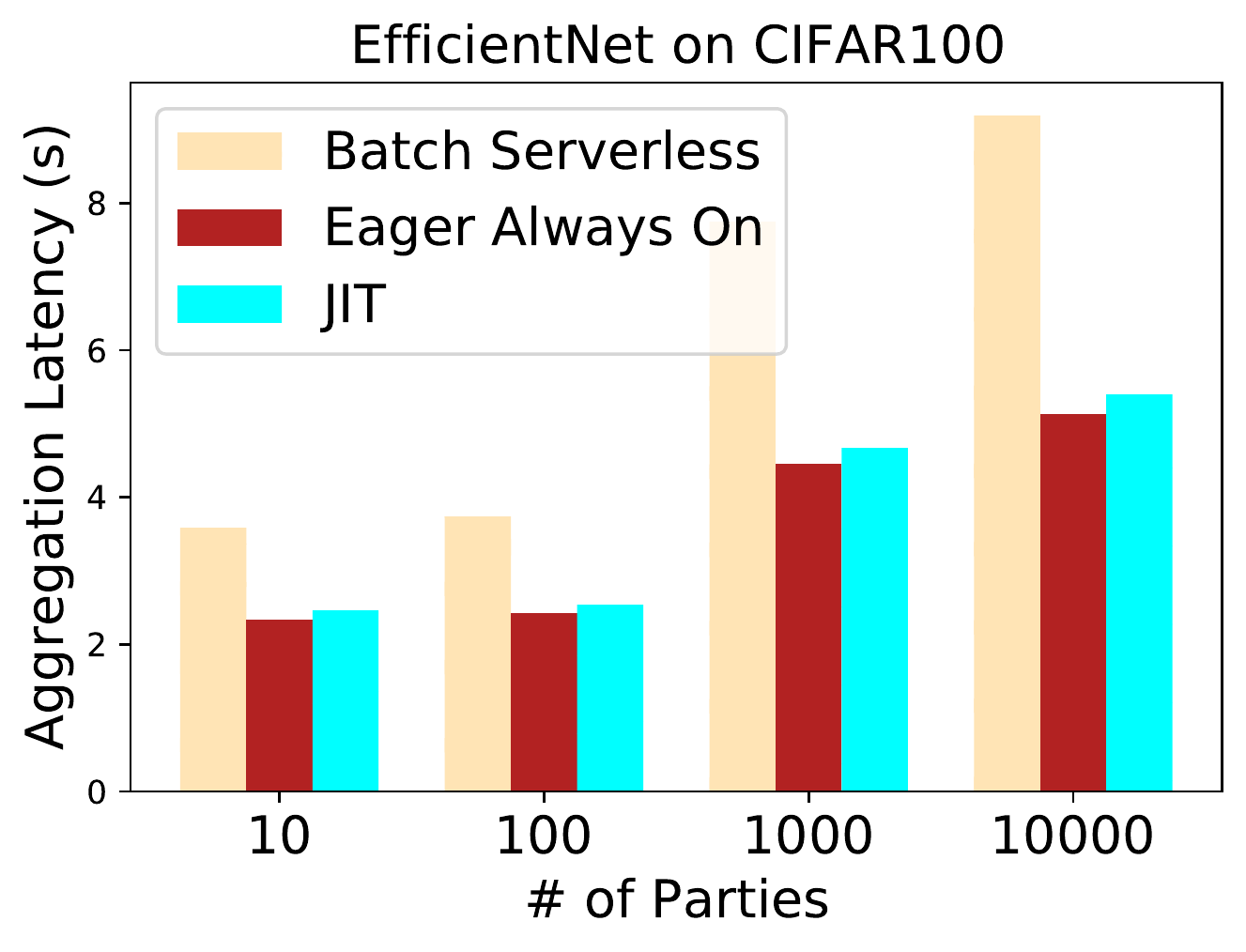}
    \includegraphics[width=0.32\textwidth]{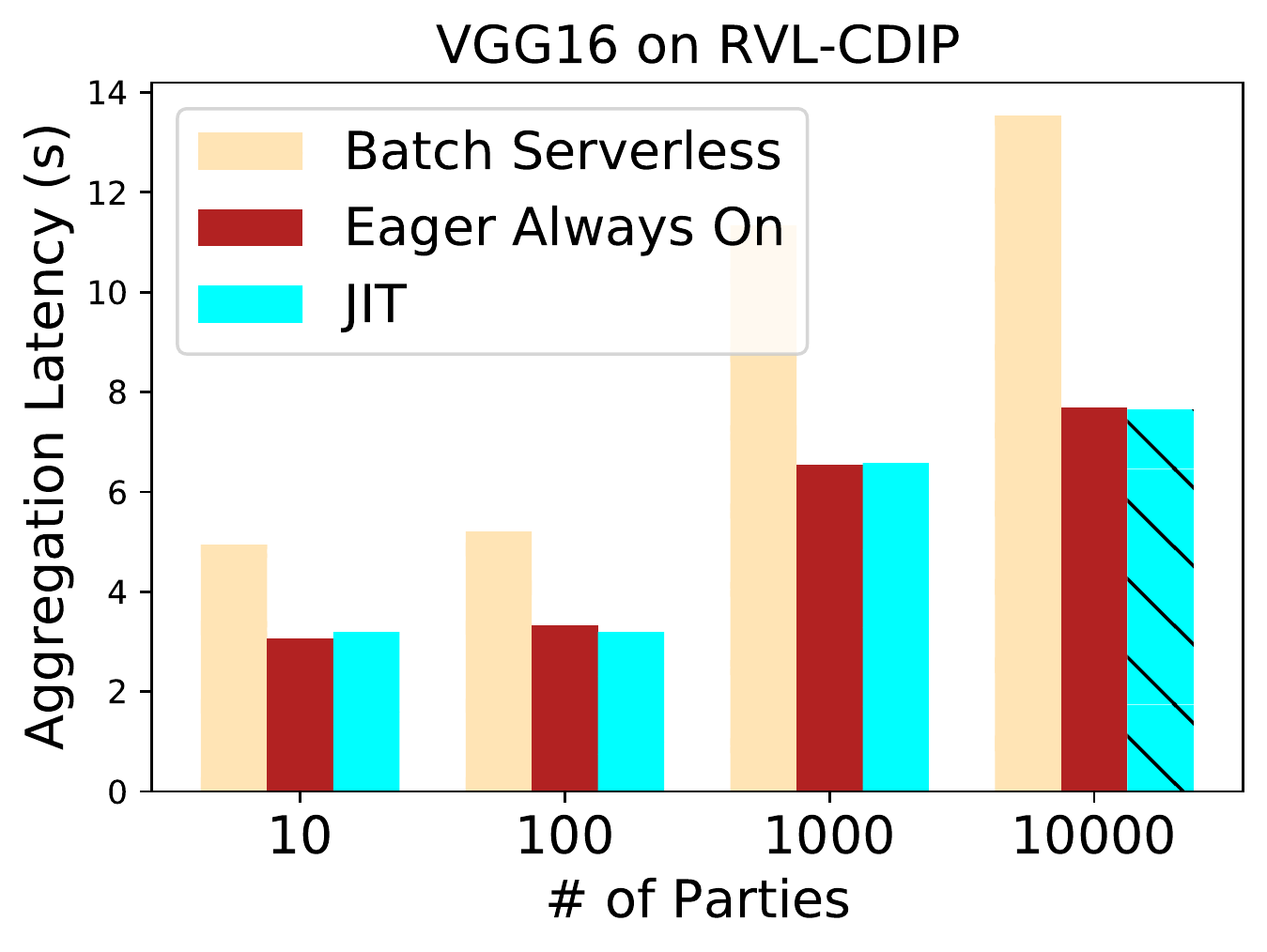}
    \includegraphics[width=0.32\textwidth]{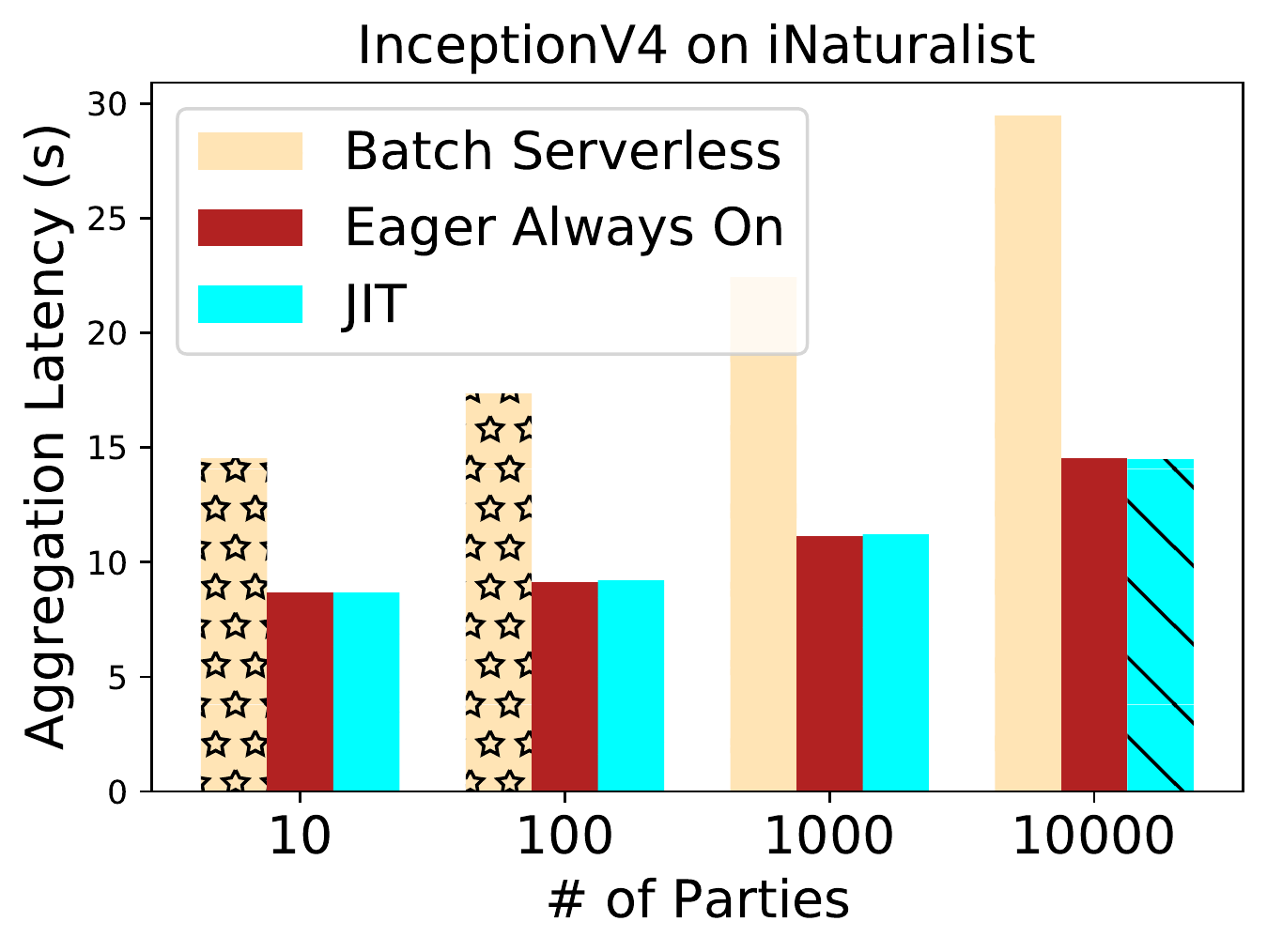}
    \caption{Aggregation Latency (s) -- time taken for aggregation to finish after the last model update is available. Heterogeneous intermittent parties}
    \label{fig:agglatency-int}
\end{figure*}

\begin{figure*}[htb]
    \centering
    \includegraphics[width=0.32\textwidth]{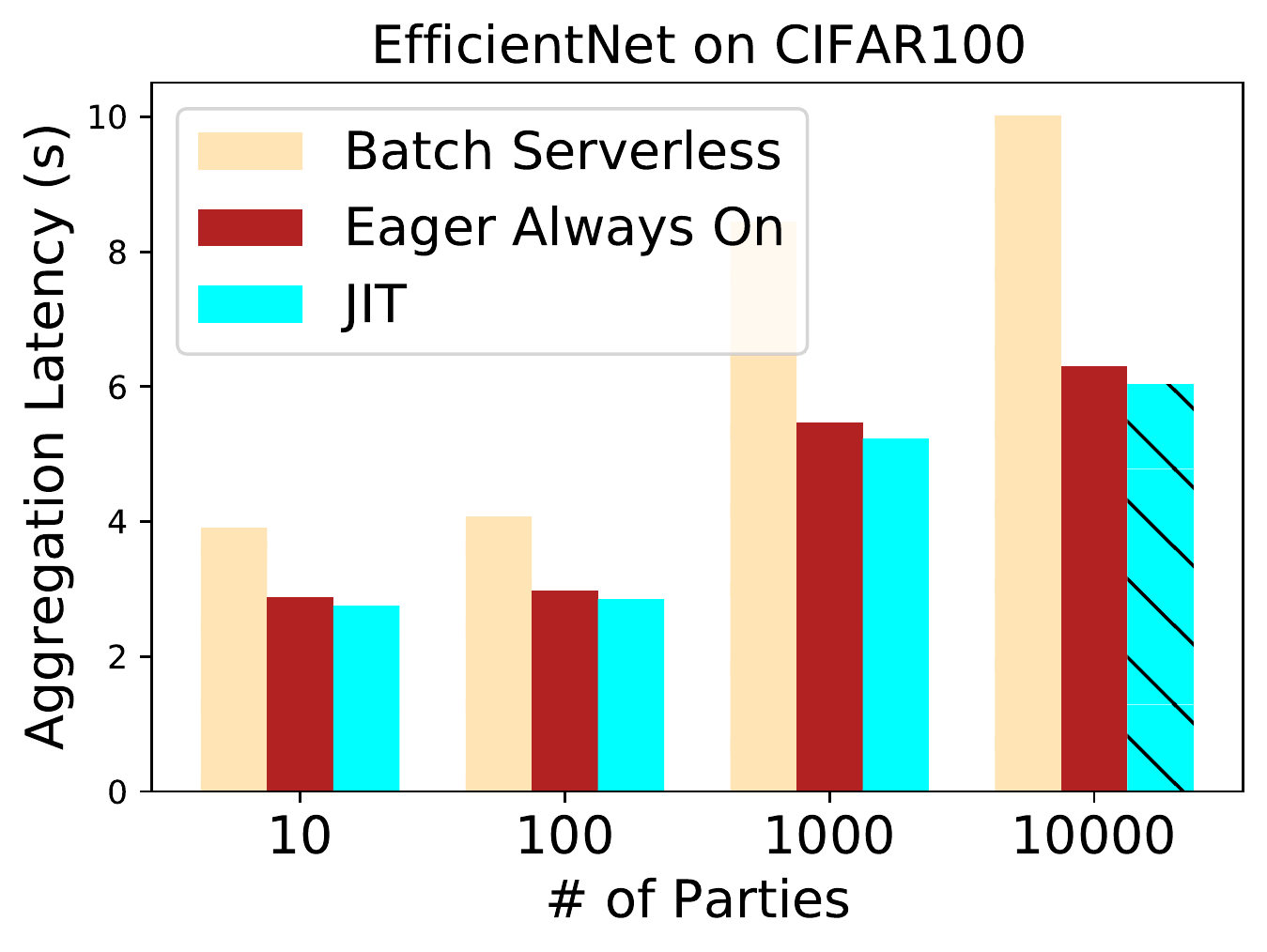}
    \includegraphics[width=0.32\textwidth]{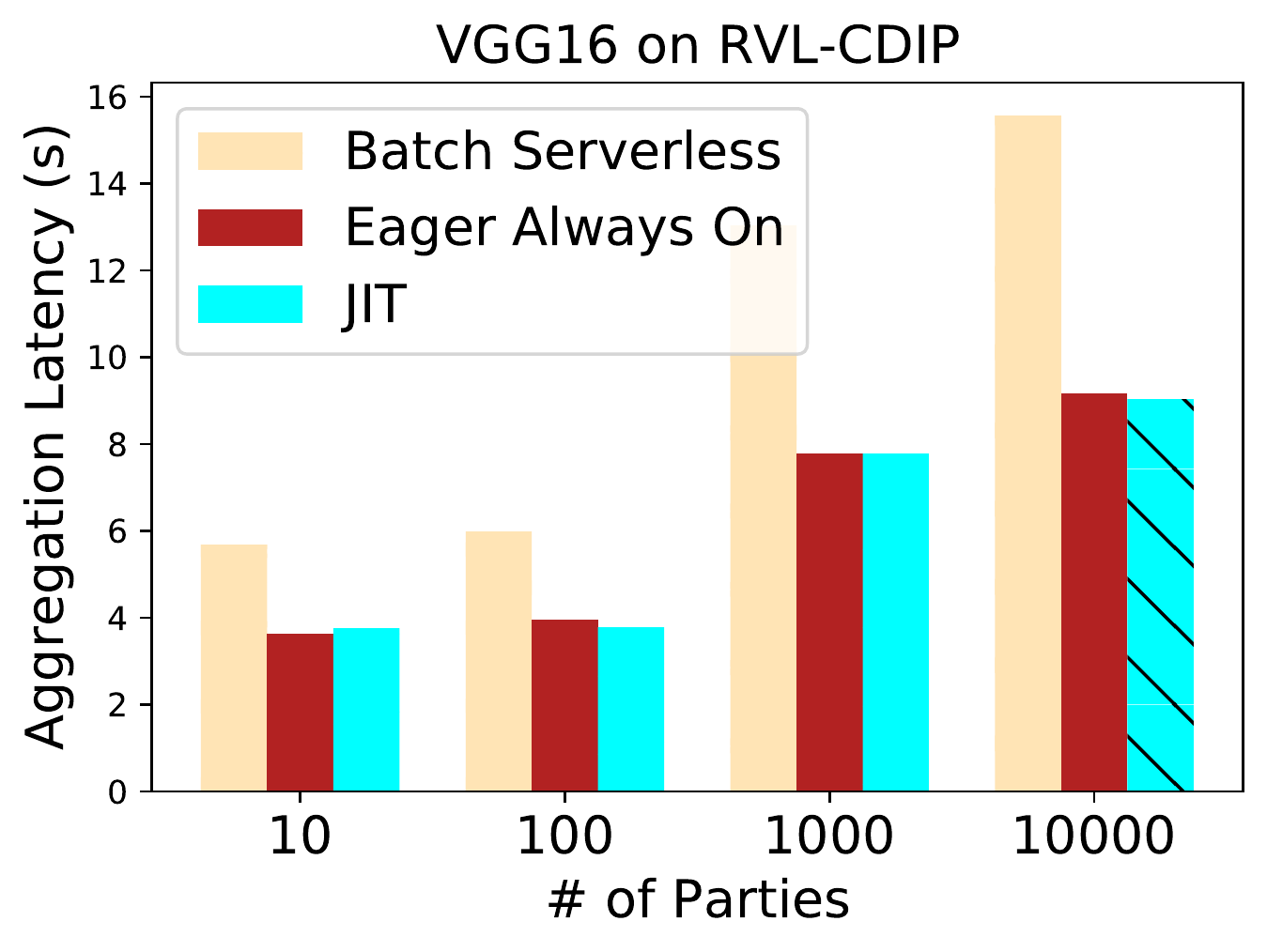}
    \includegraphics[width=0.32\textwidth]{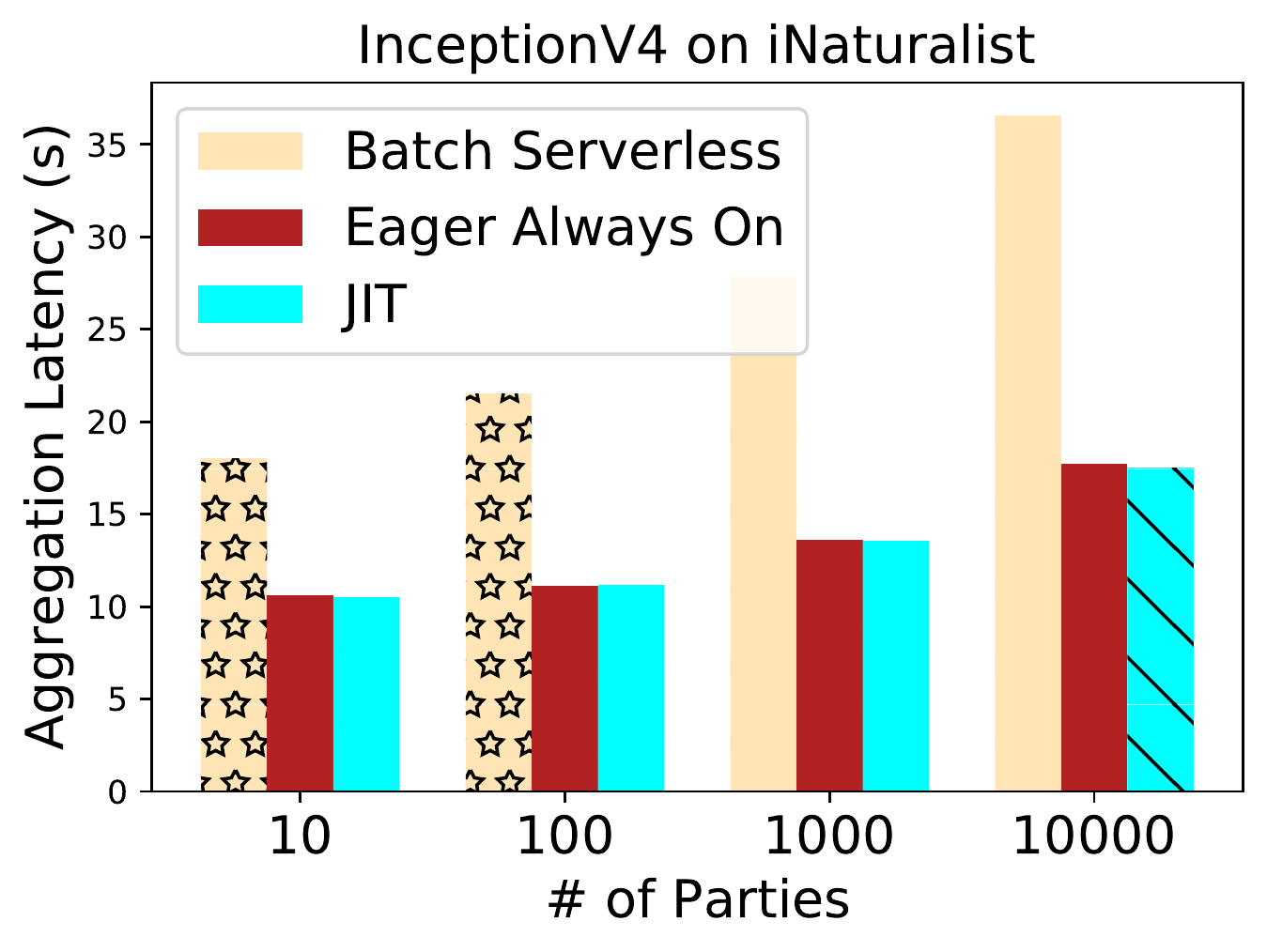}
    \caption{Aggregation Latency (s) -- time taken for aggregation to finish after the last model update is available. Heterogeneous active parties}
    \label{fig:agglatency-act}
\end{figure*}

\begin{figure*}[htbp]
    \centering
    \setlength{\tabcolsep}{0.5em}
\begin{tabular}{|r|r|r|r|r|r|r|r|r|r|r|r|}
\toprule
\multicolumn{1}{|c|}{} & \multicolumn{4}{|c|}{Total container seconds}  & \multicolumn{4}{|c|}{Proj. Total cost US\$} & \multicolumn{3}{|c|}{Cost Savings (\%)}  \\
\multicolumn{1}{|c|}{\# Parties} &
  \multicolumn{1}{|c|}{JIT} &
  \multicolumn{1}{|c|}{Batch } &
  \multicolumn{1}{|c|}{Eager} &
  \multicolumn{1}{|c|}{Eager} &  
  \multicolumn{1}{|c|}{JIT} &
  \multicolumn{1}{|c|}{Batch} &
  \multicolumn{1}{|c|}{Eager} &
  \multicolumn{1}{|c|}{Eager} &
  \multicolumn{1}{|c|}{JIT vs.} &
  \multicolumn{1}{|c|}{JIT vs.} & 
    \multicolumn{1}{|c|}{JIT vs.} \\
 \multicolumn{1}{|c|}{} &
  \multicolumn{1}{|c|}{} &
  \multicolumn{1}{|c|}{$\lambda$} &
  \multicolumn{1}{|c|}{$\lambda$} &
  \multicolumn{1}{|c|}{AO} &  
  \multicolumn{1}{|c|}{} &
  \multicolumn{1}{|c|}{$\lambda$} &
  \multicolumn{1}{|c|}{$\lambda$} &
  \multicolumn{1}{|c|}{AO} &
  \multicolumn{1}{|c|}{Batch $\lambda$} &
  \multicolumn{1}{|c|}{Eager $\lambda$} &
    \multicolumn{1}{|c|}{Eager AO}\\
  \midrule
 \multicolumn{12}{|c|}{\bf EfficientNet-B7 on CIFAR100 using FedProx aggregation algorithm. Active homogeneous Parties.} \\
 \midrule
10    & 179   & 274   & 524   & 1723 & 0.05 & 0.07 & 0.14  & 0.46 & 28.57\% & 64.29\% & 91.67\%   \\
100   & 229   & 361   & 743   & 2653 & 0.06 & 0.1  & 0.2   & 0.71 & 40\%    & 70\% & 90.54\%     \\
1000  & 2017  & 2988  & 5691  & 22340 & 0.54 & 0.8  & 1.53  & 6.01 & 32.5\%  & 64.71\% & 90.71\%   \\
10000 & 24940 & 40860 & 78093 & 298900 & 6.71 & 11   & 21.02 & 80.46 & 39\%    & 68.08\% & 91.94\%  \\ 
\bottomrule 
 \multicolumn{12}{|c|}{\bf VGG16 on RVL-CDIP using FedSGD aggregation algorithm. Active homogeneous Parties.} \\
  \midrule
10    & 134   & 205   & 456  & 1953 & 0.04 & 0.06 & 0.12 & 0.53 & 33.33\% & 66.67\% & 94.55\% \\
100   & 170   & 244   & 523  & 3078 & 0.05 & 0.07 & 0.14  & 0.83 & 28.57\% & 64.29\% & 95.24\% \\
1000  & 1324  & 2026  & 4099 & 25250 & 0.36 & 0.55 & 1.1  & 6.8 & 34.55\% & 67.27\% & 95.15\% \\
10000 & 17298 & 30321 & 66083 & 337830 & 4.66 & 8.16 & 17.79 & 90.94 & 42.89\% & 73.81\% & 94.79\% \\  
\bottomrule
 \multicolumn{12}{|c|}{\bf InceptionV4 on iNaturalist using FedProx aggregation algorithm. Active homogeneous Parties.} \\
  \midrule
10   & 223   & 369   & 747    & 2365 & 0.06  & 0.1   & 0.2   & 0.64 & 40\%    & 70\%   & 88.68\% \\
100  & 397   & 646   & 1338   & 3354 & 0.11  & 0.17  & 0.36  & 0.9 & 35.29\% & 69.44\% & 89.69\% \\
1000 & 2940  & 5246  & 11916  & 30545 & 0.79  & 1.41  & 3.21  & 8.22 & 43.97\% & 75.39\% & 90.4\% \\
10000 & 41192 & 71048 & 162650 & 420870 & 11.09 & 19.13 & 43.79 & 113.3 & 42.03\% & 74.67\% &89.88\% \\ \bottomrule
 \multicolumn{12}{|c|}{\bf EfficientNet-B7 on CIFAR100 using FedProx aggregation algorithm. Active heterogeneous Parties.} \\

  \midrule
10    & 129   & 271   & 508   & 1767 & 0.03 & 0.07 & 0.14 & 0.48 & 57.14\% & 78.57\% & 91.49\% \\
100   & 193   & 390   & 776   & 2728 & 0.05 & 0.1  & 0.21 & 0.73 & 50\%    & 76.19\% & 91.78\% \\
1000  & 1665  & 3000  & 6083  & 22421 & 0.45 & 0.81 & 1.64 & 6.04 & 44.44\% & 72.56\% & 93.02\% \\
10000 & 21268 & 40864 & 81354 & 298965 & 5.73 & 11   & 21.9 & 80.48 & 47.91\% & 73.84\% & 92.21\% \\ \bottomrule
 \multicolumn{12}{|c|}{\bf VGG16 on RVL-CDIP using FedSGD aggregation algorithm. Active heterogeneous Parties.} \\

  \midrule
10    & 96    & 195   & 388   & 1975 & 0.03 & 0.05 & 0.1   & 0.53 & 40\%    & 70\%   & 96.36\%  \\
100   & 117   & 247   & 497   & 3123 & 0.03 & 0.07 & 0.13  & 0.84 & 57.14\% & 76.92\% & 96.47\% \\
1000  & 929   & 2036  & 4069  & 25335 & 0.25 & 0.55 & 1.1   & 6.82 & 54.55\% & 77.27\% & 96.48\% \\
10000 & 14873 & 30329 & 63099 & 337856 & 4    & 8.16 & 16.99 & 90.95 & 50.98\% & 76.46\% & 95.87\% \\ \bottomrule
 \multicolumn{12}{|c|}{\bf InceptionV4 on iNaturalist using FedProx aggregation algorithm. Active heterogeneous Parties.} \\

  \midrule
10   & 175   & 371   & 869    & 2056 & 0.05  & 0.1   & 0.23  & 0.55 & 50\%    & 78.26\% & 90.91\% \\
100  & 285   & 623   & 1340   & 3616 & 0.08  & 0.17  & 0.36  & 0.97 & 52.94\% & 77.78\% & 91.67\% \\
1000 & 2719  & 5252  & 11153  & 30630 & 0.73  & 1.41  & 3     & 8.25 & 48.23\% & 75.67\% & 91.25\% \\
10000 & 39058 & 71027 & 152020 & 420903 & 10.51 & 19.12 & 40.92 & 113.31 & 45.03\% & 74.32\% & 91.56\% \\ \bottomrule
\multicolumn{12}{|c|}{\bf EfficientNet-B7 on CIFAR100 using FedProx aggregation algorithm. Intermittent heterogeneous Parties.} \\
  \midrule
10    & 201   & 282   & 380  & 634 & 0.05 & 0.08 & 0.1  & 0.17 & 28.72\% & 47.11\% & $>$ 99\% \\
100   & 306   & 460   & 654  & 576 & 0.08 & 0.12 & 0.18 & 0.16 & 33.48\% & 53.21\% & $>$ 99\% \\
1000  & 801   & 1289  & 2426 & 10516 & 0.22 & 0.35 & 0.65 & 2.83 & 37.86\% & 66.98\% & $>$ 99\% \\
10000 & 13102 & 20786 & 49023 & 105021 & 3.53 & 5.6  & 13.2 & 28.27 & 36.97\% & 73.27\% & $>$ 99\% \\ \bottomrule
 \multicolumn{12}{|c|}{\bf VGG16 on RVL-CDIP using FedSGD aggregation algorithm. Intermittent heterogeneous Parties.} \\
  \midrule
10    & 235   & 283   & 392   & 33043    & 0.06 & 0.08 & 0.11  & 8.9    & 16.96\% & 40.05\% & $>$ 99\% \\
100   & 280   & 418   & 576   & 33037    & 0.08 & 0.11 & 0.16  & 8.89     & 33.01\% & 51.39\% & $>$ 99\% \\
1000  & 1666  & 2853  & 4752  & 510039    & 0.45 & 0.77 & 1.28  & 137.3    & 41.61\% & 64.94\% & $>$ 99\% \\
10000 & 19474 & 35419 & 72258 & 5700030    & 5.24 & 9.53 & 19.45 & 1534.45    & 45.02\% & 73.05\% & $>$ 99\% \\ \bottomrule
\multicolumn{12}{|c|}{\bf InceptionV4 on iNaturalist using FedProx aggregation algorithm. Intermittent heterogeneous Parties.} \\
  \midrule
10   & 420    & 534    & 794    & 34365 & 0.11  & 0.14  & 0.21  & 9.25 & 21.35\% & 47.1\% & $>$ 99\%  \\
100  & 417    & 663    & 840    & 34358 & 0.11  & 0.18  & 0.23  & 9.25 & 37.1\%  & 50.36\% & $>$ 99\% \\
1000 & 11618  & 17672  & 30434  & 734456 & 3.13  & 4.76  & 8.19  & 197.72 & 34.26\% & 61.83\% & $>$ 99\% \\
10000 & 138469 & 232010 & 520420 & 6783036 & 37.28 & 62.46 & 140.1 & 1825.99 & 40.32\% & 73.39\% & $>$ 99\% \\ \bottomrule
\end{tabular}
    \caption{Resource usage and projected cost, using container cost/s of 0.0002692 US\$ (source Microsoft Azure\cite{azurepricing}). $\lambda$ means serverless and AO means ``Always-On''}~\label{tbl:master}
\end{figure*}

In this section, we evaluate the efficacy of JIT aggregation, by comparing it to
eager aggregation, and batched eager aggregation. Specifically, we evaluate the 
(i) \emph{efficiency} by examining whether JIT aggregation increases the latency
of an FL job, as perceived by a participant, (ii) \emph{scalability} by examining the
impact of the number of parties on latency, and (iii) \emph{resource efficiency}, by measuring 
resource consumption (in terms 
of the number and duration of containers used for aggregation) and projected total cost.

\subsection{Implementation \& Experimental Setup}

For Eager ``Always-On'' aggregation, we simply use IBM FL~\cite{ibmfl}.
For Eager Serverless (Eager $\lambda$), we take the aggregation code from IBM FL~\cite{ibmfl}
and execute it in parallel using the serverless computing feature of the 
Ray distributed computing platform. We employ Ray (as opposed to KNative and Openwhisk) because of
its native support for Python (and consequently ML frameworks like Tensorflow and Pytorch).
Batched Serverless is a variant of Eager Serverless where aggregation is triggered after 
batches of model updates have been sent and are available at the message queue; we implement
Batched Serverless and our JIT strategy using Ray as well. Aggregation was executed on a Kubernetes cluster on CPUs, using Ray on Docker containers. 
Each container (with a Ray executor) was equipped with 2 vCPUs (2.2 Ghz, Intel Xeon 4210) and 
4 GB RAM. Parties were emulated, and distributed over
four datacenters (different from the aggregation datacenter) to emulate geographic distribution.
Each party was also executed inside Docker containers (2 vCPUs and 8 GB RAM) on Kubernetes, and these containers
had dedicated resources. We actually had parties running training to emulate realistic federated
learning, as opposed to using, e.g., Tensorflow Federated simulator.

\subsection{Metrics}

We execute Ray serverless functions using 
Docker containers on Kubernetes pods in our datacenter, and measure the number of \emph{container seconds}
used by an FL job from start to finish. Container seconds is calculated by multiplying the number of 
containers used with the time that each container was used/alive. This includes all the resources used 
by the ancillary services, 
including MongoDB (for metadata), Kafka and Cloud Object Store. Measuring \emph{container seconds} helps us use
publicly available pricing from cloud providers like Microsoft Azure to project the monetary cost
of aggregation, in both cases, and project cost savings.

Since our JIT strategy defers aggregation as much
as possible, overheads in our work will usually manifest in the form of increased \emph{aggregation latency}.
Given that aggregation depends on whether the expected number of model updates are available, we
define \emph{aggregation latency} as the time elapsed between the reception of the last model update
and the availability of the aggregated/fused model. It is measured for each FL synchronization round,
and the reported numbers in the paper are averaged over all the rounds of the FL job. We want aggregation latency to be as
low as possible. Scalability, or the lack thereof, of any FL aggregation architecture, also manifests in the form
of increased aggregation latency when the number of parties rises.

\subsection{Workloads}

We select three real-world federated learning jobs -- 
two image classification tasks from the Tensorflow Federated (TFF)~\cite{tff-benchmark} benchmark
and one popular document classification task. From TFF~\cite{tff-benchmark}, we select (i) CIFAR100 dataset which can be 
distributed over 10-10000 parties, with classification performed using the EfficientNet-B7 model and the FedProx~\cite{fedprox}
aggregation algorithm and (ii) iNaturalist dataset
which can be distributed over 10-10000 parties, with classification performed using the InceptionV4
model and FedProx~\cite{fedprox} aggregation algorithm. Thus, 
we consider two types of images and two models of varying sizes. We do not consider other workloads
from TFF because they involve less than 1000 parties. For additional diversity, we consider a third workload
using the VGG16~\cite{vgg16-rvlcdip} model and FedSGD~\cite{bonawitz2019towards} aggrgeation algorithm on RVL-CDIP~\cite{rvlcdip} document classification dataset. Each job was executed for 50 synchronization rounds, with model fusion happening after every local epoch.
For all scenarios, the datasets were partitioned in a realistic non-IID manner. For batched aggregation, aggregation was 
triggered every (2,10,100,100) model updates for the (10, 100, 1000, 10000) party scenarios.

In the case of active participants, model training and update is straightforward. To emulate intermittent 
participants, we used a random update scheme -- within the time interval allotted to an FL round, each participant
would send their model update at a random time. In the case of homogeneous parties, each party was allotted
2 vCPUs and 4GB RAM with a equal slice of the dataset chosen in a non-IID manner. That is, each party got the same 
amount of data but the distribution of the data among the labels was different among parties. 
For heterogeneous parties, each party was randomly allotted 1 or 2 vCPUs with (2, 4, 6, 8) GB of RAM, also chosen
randomly.

\subsection{Aggregation Latency}

First, we examine the impact of JIT aggregation on the latency of the FL job. Figures~\ref{fig:agglatency-act} and \ref{fig:agglatency-int} illustrate the effect of JIT aggregation on latency, for heterogeneous parties
with and without active participation. The results for homogeneous parties is very similar; we omit these due to space constraints. We observe that the perceived effect of JIT aggregation (as measured by latency from the parties' side)
is negligible when compared to eager aggregation. This is a validation of our central thesis
that training time can be accurately estimated in FL. Once this is done, and aggregation scheduled in time for the final update
from the parties, there is no impact on latency. This is true in the case of heterogeneous parties as well, 
whether the training time is estimated directly from minibatch/epoch time provided by the parties or using 
linear regression. The aggregation latency of batched aggregation is generally higher than that of eager or JIT schemes,
because batching keeps waiting for certain amounts of updates to arrive, and in cases where updates are bunched up due
to heterogeneity, completing aggregation takes additional time. We also observe that while increasing the number
of parties slightly increases overall latency, JIT aggregation continues to perform as well as eager aggregation.

\subsection{Resource Efficiency}

Next, we examine the resource and cost savings realized by deferring aggregation. Figure~\ref{tbl:master} illustrates the resource savings with active homogeneous, active heterogeneous and intermittent 
heterogeneous parties for all three model/dataset combinations.
We observe that, eager always-on (Eager AO) aggregation in existing FL platforms is the most resource intensive,
irrespective of whether parties are active or intermittent, homogeneous or heterogeneous.
Eager serverless (Eager $\lambda$) performs better than Eager AO, because it involves dynamically
deploying aggregators and relinquishes resources when possible. Batched serverless (Batch $\lambda$) 
further improves utilization because it reduces the number of times the aggregator has to be deployed,
and it also ensures that each deployed aggregator has substantial work -- Eager $\lambda$ may deploy an aggregator
to process one or two model updates, while Batch $\lambda$ ensures at least a batch of updates to process,
amortizing deployment overheads and context switches.

Figure~\ref{tbl:master} starts with active, homogeneous parties. This is the ideal case for 
training time estimation, 
and we observe a healthy 60-75\% resource and cost savings with respect to eager serverless aggregation
and $\approx$ 90\% with respect to eager always-on (IBM FL).
In batched aggregation, aggregators are not deployed as often as the eager strategy. Hence, the savings with 
respect to batched aggregation is not as high as the eager case, but nevertheless significant at 28-40\%.
As the number of parties increases, the overall resource usage increases significantly for all experiments,
but the savings persist. Thus, JIT aggregation has the same or better latency than eager and batched aggregation
but saves a large chunk of datacenter resources, which can be used by other FL jobs and other workloads.

We observe from Fig.~\ref{tbl:master} that this trend persists for active, heterogeneous entities, 
where training time can still be predicted with
high accuracy. Resource savings w.r.t Eager $\lambda$ and Batch $\lambda$ are higher
than active homogeneous parties because model updates arrive at different times, which makes the
JIT strategy more useful. The case with intermittent participants who are heterogeneous
 represents a challenging case for JIT aggregation, because updates arrive any time during the aggregation
window and these experiments test our priority setting strategy of Section~\ref{sec:priority}. Aggregation
latency continues to remain low (in Figure~\ref{tbl:master}) while still achieving savings of 70+\% with 
respect to eager aggregation.

\section{Related Work}~\label{related}

To the best of our knowledge, our work is the first to explore lazy or deferred
aggregation for federated learning. A broad overview of the area of federated learning 
is beyond the scope of this paper; for that, we 
refer the reader to \cite{kairouz2019advances, yang-floverview, berkeley-serverless}. Scalable and
efficient aggregation is a key 
problem in federated learning, as identified by \cite{kairouz2019advances, ibmfl, fate}.
While \cite{bonawitz2019towards}
uses hierarchical aggregation, its programming model is different from our work. Its primary goal is 
scalability and consequently, it deploys long lived actors and 
seems to implement the eager aggregation model. Oort~\cite{oort} is another recent system that
prioritizes the subset of clients who have both data that offers the greatest utility in 
improving model accuracy and the compute to run training quickly. But Oort does not
address the challenge of scheduling aggregation effectively and providing aggregation as a cloud service.

A number of ML frameworks -- Siren~\cite{siren}, Cirrus~\cite{cirrus} and
the work by LambdaML~\cite{jiang-serverless-ml} use serverless functions
for centralized (not federated) ML and DL training.
Siren~\cite{siren} allows users to train models (ML, DL and RL) in the cloud 
using serverless functions with the goal to reduce programmer burden involved
in using traditional ML frameworks and cluster management technologies for
large scale ML jobs. It also contains optimization algorithms to tune training
performance and reduce training cost using serverless functions. 
Cirrus~\cite{cirrus} goes further, supporting end-to-end centralized ML training workflows
and hyperparameter tuning using serverless functions. 
LambdaML~\cite{jiang-serverless-ml} analyzes
the cost-performance trade-offs between IaaS and serverless
for datacenter/cloud hosted centralized ML training.
Our work differs from Siren, Cirrus and LambdaML in 
significant ways -- Distributed ML (in Siren, Cirrus and LambdaML) is different from FL. Distributed ML involves 
centralizing data at a data center or cloud service and performing training at a central location.
In contrast, with FL, data never leaves a participant. FL's privacy guarantees are much stronger
and trust requirements much lower than that of distributed ML. 
FedLess~\cite{fedless} has the ability to run a single eager 
aggregator as a cloud function, but does not
have the ability to parallelize aggregation.

\section{Conclusions and Future Work}~\label{conclusions}

In this paper, we take a fresh look at the problem of scalable aggregation for 
federated learning. While FL has been increasingly adopted, existing research has 
gaps in addressing how cloud providers should manage large numbers of 
FL jobs if they decide to become a nexus between their customers and offer
FL-as-a-service. We demonstrate that using a JIT strategy to defer aggregation
until the point it is needed can be helpful and resource efficient, with negligible
overheads. JIT aggregation is applicable to a variety of different scenarios,
whether aggregators are deployed as serverless functions or containers, with or without
cluster management systems. It also works with multiple existing aggregation algorithms, as demonstrated 
in our empirical evaluation.


\begin{thebibliography}{10}
\providecommand{\url}[1]{#1}
\csname url@samestyle\endcsname
\providecommand{\newblock}{\relax}
\providecommand{\bibinfo}[2]{#2}
\providecommand{\BIBentrySTDinterwordspacing}{\spaceskip=0pt\relax}
\providecommand{\BIBentryALTinterwordstretchfactor}{4}
\providecommand{\BIBentryALTinterwordspacing}{\spaceskip=\fontdimen2\font plus
\BIBentryALTinterwordstretchfactor\fontdimen3\font minus
  \fontdimen4\font\relax}
\providecommand{\BIBforeignlanguage}[2]{{%
\expandafter\ifx\csname l@#1\endcsname\relax
\typeout{** WARNING: IEEEtran.bst: No hyphenation pattern has been}%
\typeout{** loaded for the language `#1'. Using the pattern for}%
\typeout{** the default language instead.}%
\else
\language=\csname l@#1\endcsname
\fi
#2}}
\providecommand{\BIBdecl}{\relax}
\BIBdecl

\bibitem{kairouz2019advances}
P.~Kairouz, H.~B. McMahan, B.~Avent, A.~Bellet, M.~Bennis, A.~N. Bhagoji,
  K.~Bonawitz, Z.~Charles, G.~Cormode, R.~Cummings \emph{et~al.}, ``Advances
  and Open Problems in Federated Learning,'' \emph{arXiv preprint
  arXiv:1912.04977}, 2019.

\bibitem{bonawitz2019towards}
K.~Bonawitz, H.~Eichner, W.~Grieskamp, D.~Huba, A.~Ingerman, V.~Ivanov,
  C.~Kiddon, J.~Kone{\v{c}}n{\`y}, S.~Mazzocchi, H.~B. McMahan \emph{et~al.},
  ``Towards Federated Learning at Scale: System Design,'' \emph{arXiv preprint
  arXiv:1902.01046}, 2019.

\bibitem{nvidia-covid}
I.~Dayan et. al., ``Federated Learning for Predicting Clinical Outcomes in
  Patients with COVID-19,'' \emph{Nature Medicine}, 2021.

\bibitem{ibmflpublic}
IBM~Corporation, ``IBM Federated Learning Library'', \url{https://github.com/IBM/federated-learning-lib}, 2021.

\bibitem{ibmfl}
H.~Ludwig, N.~Baracaldo, G.~Thomas, Y.~Zhou, A.~Anwar, S.~Rajamoni, Y.~Ong,
  J.~Radhakrishnan, A.~Verma, M.~Sinn, M.~Purcell, A.~Rawat, T.~Minh,
  N.~Holohan, S.~Chakraborty, S.~Whitherspoon, D.~Steuer, L.~Wynter, H.~Hassan,
  S.~Laguna, M.~Yurochkin, M.~Agarwal, E.~Chuba, and A.~Abay, ``IBM Federated
  Learning: an Enterprise Framework White Paper v0.1,'' \emph{arXiv preprint
  arXiv:2007.10987}, 2020.

\bibitem{fate}
\BIBentryALTinterwordspacing
Y.~Liu, T.~Fan, T.~Chen, Q.~Xu, and Q.~Yang, ``FATE: An Industrial Grade
  Platform for Collaborative Learning with Data Protection,'' \emph{Journal of
  Machine Learning Research}, vol.~22, no. 226, pp. 1--6, 2021. [Online].
  Available: \url{http://jmlr.org/papers/v22/20-815.html}
\BIBentrySTDinterwordspacing

\bibitem{nvflare}
NVIDIA, ``NVIDIA Federated Learning Application Runtime Environment,''
  \url{https://github.com/NVIDIA/NVFlare}, 2021.

\bibitem{azurepricing}
Microsoft~Corporation, ``{Azure Container Instances pricing},''
  \url{https://azure.microsoft.com/en-us/pricing/details/container-instances/},
  2021.

\bibitem{tff-benchmark}
Tensorflow~Project, ``Using TFF for Federated Learning Research,''
  \url{https://www.tensorflow.org/federated}, 2022.

\bibitem{fedprox}
T.~Li, A.~K. Sahu, M.~Zaheer, M.~Sanjabi, A.~Talwalkar, and V.~Smith,
  ``Federated Optimization in Heterogeneous Networks,'' in \emph{Conference on Machine Learning and Systems (MLSys)},
  I.~Dhillon, D.~Papailiopoulos, and V.~Sze, Eds., 2020, pp. 429--450.

\bibitem{vgg16-rvlcdip}
A.~Das, S.~Roy, U.~Bhattacharya, and S.~K. Parui, ``Document Image
  Classification with Intra-domain Transfer Learning and Stacked Generalization
  of deep convolutional neural networks,''[Online]. Available: \url{https://arxiv.org/abs/1801.09321}, 2018.

\bibitem{rvlcdip}
A.~W. Harley, A.~Ufkes, and K.~G. Derpanis, ``Evaluation of Deep Convolutional
  Nets for Document Image Classification and Retrieval,'' in
  \emph{International Conference on Document Analysis and Recognition}.\hskip
  1em plus 0.5em minus 0.4em\relax IEEE, 2015, pp. 991--995.

\bibitem{yang-floverview}
\BIBentryALTinterwordspacing
Q.~Yang, Y.~Liu, T.~Chen, and Y.~Tong, ``Federated Machine Learning: Concept
  and Applications,'' \emph{ACM Trans. Intell. Syst. Technol.}, vol.~10, no.~2,
  jan 2019. [Online]. Available: \url{https://doi.org/10.1145/3298981}
\BIBentrySTDinterwordspacing

\bibitem{berkeley-serverless}
E.~Jonas, J.~Schleier-Smith, V.~Sreekanti, C.-C. Tsai, A.~Khandelwal, Q.~Pu,
  V.~Shankar, J.~Carreira, K.~Krauth, N.~Yadwadkar, J.~E. Gonzalez, R.~A. Popa,
  I.~Stoica, and D.~A. Patterson, ``Cloud Programming Simplified: A Berkeley
  view on Serverless Computing,'' 2019.

\bibitem{oort}
\BIBentryALTinterwordspacing
F.~Lai, X.~Zhu, H.~V. Madhyastha, and M.~Chowdhury, ``Oort: Efficient Federated
  Learning via Guided Participant Selection,'' in \emph{15th {USENIX} Symposium
  on Operating Systems Design and Implementation ({OSDI} 21)}.\hskip 1em plus
  0.5em minus 0.4em\relax {USENIX} Association, Jul. 2021, pp. 19--35.
  [Online]. Available:
  \url{https://www.usenix.org/conference/osdi21/presentation/lai}
\BIBentrySTDinterwordspacing

\bibitem{siren}
H.~Wang, D.~Niu, and B.~Li, ``Distributed Machine Learning with a Serverless
  Architecture,'' in \emph{IEEE INFOCOM 2019 - IEEE Conference on Computer
  Communications}, 2019, pp. 1288--1296.

\bibitem{cirrus}
\BIBentryALTinterwordspacing
J.~Carreira, P.~Fonseca, A.~Tumanov, A.~Zhang, and R.~Katz, ``Cirrus: A
  Serverless Framework for End-to-End ML Workflows,'' in \emph{ACM Symposium on Cloud Computing SoCC '19}.\hskip
  1em plus 0.5em minus 0.4em\relax New York, NY, USA: ACM, 2019, p. 13–24.
  [Online]. Available: \url{https://doi.org/10.1145/3357223.3362711}
\BIBentrySTDinterwordspacing

\bibitem{jiang-serverless-ml}
J.~Jiang, S.~Gan, Y.~Liu, F.~Wang, G.~Alonso, A.~Klimovic, A.~Singla, W.~Wu,
  and C.~Zhang, ``Towards Demystifying Serverless Machine Learning Training,''
  in \emph{ACM SIGMOD}, 2021.

\bibitem{fedless}
A.~Grafberger, M.~Chadha, A.~Jindal, J.~Gu, and M.~Gerndt, ``Fedless: Secure
  and Scalable Federated Learning Using Serverless Computing,'' in \emph{IEEE
  International Conference on BigData}, 2021.

\end{thebibliography}
\end{document}